\documentclass[conference]{IEEEtran}
\IEEEoverridecommandlockouts
\usepackage{amsmath,amssymb,amsfonts}
\usepackage{algorithm}
\usepackage{graphicx}
\usepackage{textcomp}
\usepackage{caption}
\usepackage{xcolor}
\usepackage{caption} 
\usepackage{booktabs} 
\usepackage{tabularx}
\usepackage{biblatex}  
\usepackage{subcaption}
\usepackage{graphicx} 
\usepackage{algpseudocode}
\usepackage{amsmath}
\def\BibTeX{{\rm B\kern-.05em{\sc i\kern-.025em b}\kern-.08em
    T\kern-.1667em\lower.7ex\hbox{E}\kern-.125emX}}
\begin{document}

\title{Optimizing V2V Unicast Communication Transmission with Reinforcement Learning and Vehicle Clustering\\

}

\author{\IEEEauthorblockN{1\textsuperscript{st} Long Tan*}
\IEEEauthorblockA{\textit{Department of Computer Science and Technology} \\
\textit{Heilongjiang University} \\
Harbin,China \\
tanlong@hlju.edu.cn}
\and
\IEEEauthorblockN{2\textsuperscript{nd} Yu Wang}
\IEEEauthorblockA{\textit{Department of Computer Science and Technology} \\
\textit{Heilongjiang University}\\
Harbin, China \\
wagnyu@126.com}

}

\maketitle

\begin{abstract}
 Efficient routing algorithms based on vehicular ad hoc networks (VANETs) play an important role in emerging intelligent transportation systems. This highly dynamic topology faces a number of wireless communication service challenges. In this paper, we propose a protocol based on reinforcement learning and vehicle node clustering, the protocol is called Qucts, solve vehicle-to-fixed-destination or V2V messaging problems. Improve message delivery rates with minimal hops and latency, link stability is also taken into account. The agreement is divided into three levels, first cluster the vehicles, each cluster head broadcasts its own coordinates and speed, to get more cluster members. Also when a cluster member receives another cluster head broadcast message, the cluster head generates a list of surrounding clusters, find the best cluster to the destination as the next cluster during message passing. Second, the protocol constructs a Q-value table based on the state after clustering, used to participate in the selection of messaging clusters. Finally, we introduce parameters that express the stability of the vehicle within the cluster, for communication node selection. This protocol hierarchy makes Qucts an offline and online solution. In order to distinguish unstable nodes within a cluster, Coding of each road, will have vehicles with planned routes, For example, car hailing and public bus. Compare the overlap with other planned paths vehicles in the cluster, low overlap is labeled as unstable nodes. Vehicle path overlap rate without a planned path is set to the mean value. Comparing Qucts with existing routing protocols through simulation, Our proposed Qucts scheme provides large improvements in both data delivery rate and end-to-end delay reduction.
\end{abstract}


\section{Introduction}
VANET is a key component of an intelligent transportation system, it includes vehicle-to-vehicle (V2V) communication, vehicle and infrastructure (V2I) communications. V2I performance depends on the coverage of the roadside unit (RSU). V2I communications utilize infrastructure located at the roadside, provide internet access for vehicles, achieving wide-scale information dissemination. In comparison, V2V communication only allows the exchange of information between neighboring vehicles. However, V2V communication has two main advantages over V2I: V2V shorter communication distance reduces path loss, helps improve transmission reliability; V2V communications can support delay-sensitive vehicle applications, these applications occur between neighboring vehicles, for example, traffic hazard warnings.However, the disadvantage of V2V is that vehicles joining or leaving can cause frequent communication breakdowns in the network, Therefore designing efficient routing algorithms is a challenging task in VANET.

Unicast communication has a wide range of application scenarios in VANET, unicast communication is a one-to-one communication model, where data is transmitted from a single source to a single destination. It can be used for information exchange between vehicles, vehicle-to-infrastructure communications, remote diagnostics and maintenance, remote control as well as vehicle advertising and entertainment.

Unlike mobile self-organizing networks (MANET), VANET has fast vehicle node movement, frequent link breaks, short link maintenance time, characteristics of complex communication scenarios. Delay Tolerant Network (DTN) are specialized for networks with frequent link breaks, DTN are primarily used in small mobile devices, but cannot meet large scale in vehicle network performance standards. Location based vehicular routing is a common routing strategy in VANET,it uses the vehicle's location information to optimize packet transmission and routing decisions. This type of routing takes into account factors such as vehicle position and speed, to ensure efficient data transmission and link stability. HAEQR is also a routing algorithm based on reinforcement learning, the main goal of the algorithm is to find a good strategy in a collaborative decision-making problem with multiple agents, to maximize cumulative rewards. However, the algorithm makes each vehicle node as a learning environment for the agent, there are problems of slow convergence and high computational overheads.

To overcome the above problems. In this paper, we propose a new unicast communication transmission scheme Qucts,the solution is characterized by high data delivery rates and low end-to-end latency. The main contributions of this paper are as follows: 

\begin{itemize}
\item Differences from existing methods, consider different clusters as environment states for Q-learning, instead of vehicles. This reduces the state space considerably, increased convergence rate. When a relay node is out of transmission range, the stable nodes in the cluster will be reassigned as relay nodes. Instead of having to rediscover the optimal transmission path, reduced computational overhead. 
\item Coding for each road. For vehicles with planned routes, internet and public vehicles. Comparison of path overlap rate, vehicle nodes with high overlap are considered as stable nodes. The vehicle path overlap rate for which planning paths are not available is set to the mean value. Marking of unstable nodes, improved link stability.
\item Introduction of indicators of vehicle stability within clusters, cluster heads form an angular range based on the direction of the previous and next clusters. The most stable node in the range is selected as the relay node for communication. 
\end{itemize}

The rest of this paper is structured as follows. Section II reviews related work on VANET routing. Section III describes the vehicle clustering system model. Section IV describes the learning process of Q-learning and the selection of transmission relay nodes. Section V presents the simulation results. Section VI summarizes the paper and outlines our future work. 

\section{Related Work}

Location-based VANET routing algorithms aim to utilize vehicle geolocation information to support routing decisions, these algorithms can help vehicles communicate and transmit data more efficiently in mobile networks. Determine the transmission path of the packet based on the location of the destination node and the location of the neighboring nodes. One of the representatives is GPSR, GPSR is a greedy routing algorithm, it selects the nearest neighbor node to the target node as the next hop. Packets are transmitted through a series of peripheral nodes. Reference () proposes a related improved algorithm. 

OLSR proposes a protocol based on link state, broadcast link status information periodically through each node, includes connection status with its neighboring nodes, link quality and hop count information. The protocol is characterized by low overhead, fast convergence to the optimal path and support for multicast and unicast. But in a dense network, these control messages may result in higher network overheads, bandwidth and energy consumption. Especially in the VANET environment, due to the rapid movement of vehicles, can lead to excessive network overhead. LAR proposed a method to fully utilize node location information, protocols for Reducing Route Requests. LAR Advantage in Adapting to Network Topology Changes, and the design is relatively simple and easy to implement. However, since the nodes need to be position-checked, LAR may introduce some additional latency, especially if the nodes move around a lot. 

QLAODV is an improved routing protocol based on the Q-learning approach, it uses a Q-learning algorithm to infer the state of VANET environment, each vehicle maintains a Q-value table, and based on the Q-value table as the node forwarding routing table. However, when there are too many vehicle nodes it can lead to an overly large state space, and thus slower convergence. CHAQR proposes an improved algorithm based on Q-learning, guiding forwarding actions of nodes by introducing heuristic functions and delay information among nodes, accelerated convergence of learning. But it still doesn't solve the problem of an oversized state space. 

Qgrid proposes a new unicast routing protocol, by meshing the known maps, based on the known vehicles in the grid and the vehicles that are about to enter the grid, selecting the best grid using Q-learning. Form an optimal grid pathway, significantly reduced state space and accelerated convergence. However, the meshing algorithm still has many shortcomings. First the mesh is too large, intensive learning is less effective. Too small a grid division will result in fewer vehicles in the grid, increase the number of times a link is redisconnected, reduced link stability. Secondly, vehicle density is not uniformly distributed in complex urban environments. When the density of vehicles in the neighborhood is low, selected grids may not have vehicles, need to recalculate the optimal mesh, increased computational overhead. 

\section{Clustering Algorithms And Cluster Based Model}

We assume that all vehicles have GPS devices to obtain real-time information about their traveling direction, speed, geographic location, etc. While the vehicles are traveling, they periodically exchange hello packets, which contain basic real-time information about the vehicles, such as vehicle identifiers, cluster header IDs, and current status.

To improve the efficiency of the system, the hello packet transmission beacon interval (BI) is introduced, which can be adaptively adjusted according to the link lifetime and vehicle status. The node selected as the cluster head broadcasts hello messages at a certain interval to attract more cluster members to join. When members in other nearby clusters receive the broadcast message from the cluster head, they will pass the hello packets to the cluster head of the cluster they are in, thus enabling the entire cluster network to accurately perceive the location and status of other nearby clusters. 

\subsection{Status of Vehicles}\label{AA}

In our proposed clustering approach, the vehicles are categorized into five states: the initial state IS, the cluster head CH, the cluster member CM, the temporary cluster head TCH, and the cluster member CMT that participates in unicast transmission.

\begin{itemize}
\item \textbf{Initial State IS}: At the beginning, all vehicles are in the initial state, the vehicle does not belong to any cluster.
\item \textbf{Cluster Head CH}: There is only one cluster head in each cluster, establishing one-hop communication with the cluster members. Cluster header contains two critical lists: one is the list of cluster members for recording the information of cluster members, and the other is the list of surrounding clusters for recording the statement of neighboring clusters. Also, in unicast communication, CH assigns cluster members as relay nodes for communication.
\item \textbf{Cluster Member CM}: One-hop communication vehicles that act as cluster head CH; they are alternative nodes that participate in unicast data transmission or are only members of clusters with no particular responsibilities.
\item \textbf{Temporary Cluster Head TCH}: When a vehicle leaves its original cluster, or the timer expires in IS state, and the neighboring nodes around it are in CM state, it will claim itself as a temporary cluster head. Until it finds a new cluster or a vehicle with IS state joins.
\item \textbf{Cluster Member Transmission CMT}:  Cluster member is involved in data transmission. When the previous cluster sends a data transmission request, the cluster head selects the CMT node based on certain conditions to act as a relay node for data transmission.
\end{itemize}

After the vehicle's initial state IS is started, the vehicle sets a timer Tis until Tis expires. If there are no neighboring nodes or all neighboring nodes are in CM state, the vehicle switches its state to temporary cluster head TCH. When the condition CMjoin to become a cluster member is satisfied, the vehicle state will change from TCH to CM, or a TCH node has three or more TCH neighbors, which triggers the election of the cluster head and changes the vehicle state. When there is a neighbor vehicle with an IS state and the condition CMjoin to become a cluster member is satisfied, the vehicle state will change from IS to CM. When the situation CHcondition to become a cluster head is satisfied, the vehicle state will change from IS to CH. If the vehicle leaves the original cluster and does not belong to the new cluster, the state will return to the initial state IS. The specific process is shown in Fig 1. 

\begin{figure}[htbp]
    \centering
    \includegraphics[width=0.45\textwidth]{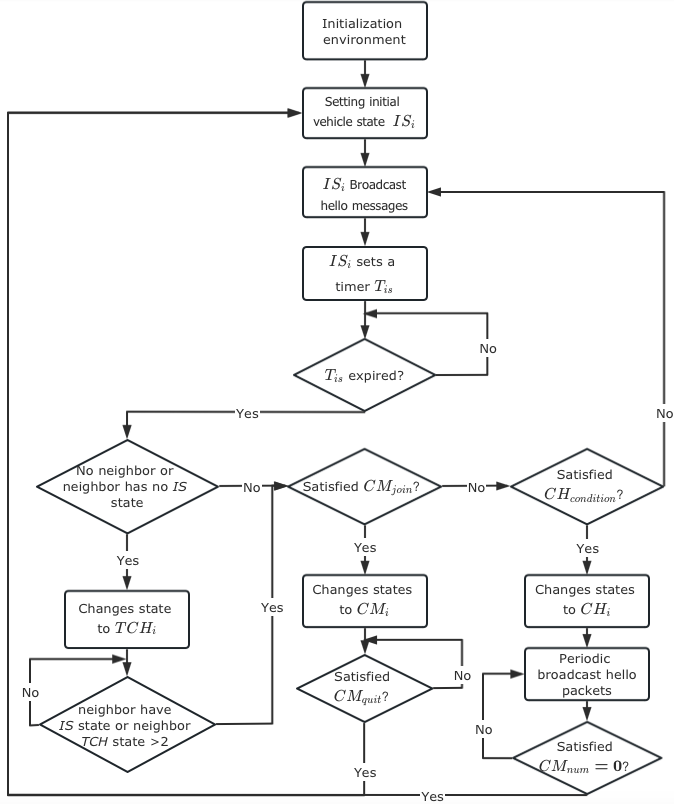}
    \caption{Vehicle status change flowchart}
    \label{fig:1}
\end{figure}

\subsection{Adaptive Tuning of Hello Packet}

In most clustering algorithms, hello packets are usually distributed according to a fixed period. The selection of the period is crucial to ensure that the information about the surrounding vehicles is updated on time, and by choosing an appropriate period, excessive communication overhead and management costs can be avoided while maintaining the real-time nature of the communication information. If the selected period is too large, the surrounding vehicles' status information may lag because of the long interval between information updates. Whereas, if too small a period is chosen, then the communication overhead and the management and maintenance costs will increase.

Therefore, we employ a state and link lifetime-based strategy called Link Expiry Time (LET)\cite{wang2013passcar} . This strategy is used to adaptively adjust the transmission period of hello packets to reflect the duration of continuous communication between two vehicles.is defined as:
\begin{equation}
    LET_{ij} = \frac{\left | \bigtriangleup v_{ij}  \right |\times TR- \bigtriangleup v_{ij}\times \bigtriangleup D_{ij}   }{\left ( \bigtriangleup v_{ij}  \right ) ^{2} } 
    \tag{1}
\end{equation}

Where TR is the distance of stable transmission between the vehicles, $\left ( x_{i},y_{i}\right ) $ is the location of the vehicle $v_{i}$, $v_{i}$ is the speed of vehicle $i$. $\bigtriangleup v_{ij} = v_{i} - v_{j}$, $\bigtriangleup D_{ij} \approx \sqrt{\left ( x_{i}- x_{j}   \right ) ^{2}+ \left ( y_{i} - y_{j}  \right ) ^{2}  } $ are the relative velocity and relative distance of $v_{i}$, $v_{j}$ respectively.

 Most vehicles are generally in the CM state in a natural road environment. Compared to cluster heads, cluster members do not need to manage other vehicles as much as cluster heads, so their Beacon Interval (BI) can be set longer. Such a setting can reduce the communication overhead while keeping the data fresh. In other words, a larger BI value can be chosen to minimize the transmission overhead when the LET is long.
 
For those CM nodes with larger LET, this indicates that they have a longer link lifetime and thus can be set with a relatively more significant BI value. However, for CHs, they usually account for less in a road environment. In time, they must connect to IS and TCH vehicles within the transmission range to get more CM nodes. Therefore, the BI value needs to be relatively small for cluster heads to maintain real-time awareness and connectivity to surrounding vehicles.

For IS state vehicles, they need to update the surrounding cluster heads' information on time so they can join the cluster as soon as possible. Similarly, TCH vehicles need to quickly sense the nearby environment to enter new clusters or to enable nearby IS-state vehicles to join and form new clusters. Therefore, the BI value also needs to be set relatively small for both cases. Based on the above description, we set the beacon interval BI as follows:
\vspace{-0.1em}
\begin{equation}
BI_{i} = \begin{cases}  & \text{ 0.3 }  \quad  ST_{i} \in \left ( CH,IS,TCH \right )  \\  & \text{ 0.3 }  \quad  LET_{ij}\in\left ( 0,0.5 \right ) ,  \ ST_{i}\in \left ( CM \right )  \\  & \text{ 0.5 }  \quad  LET_{ij}\in\left [ 5,20 \right ),  \ \ ST_{i}\in \left ( CM \right ) \\  & \text{ 1.0 }  \quad  LET_{ij}\in\left [ 20,+  \infty  \right ),  \ ST_{i}\in \left ( CM \right ) \\\end{cases}
    \tag{2}
\end{equation}

Where $ST_{i}$ is the current state of vehicle $v_{i}$, $LET_{ij}$ is the LET between CM vehicle $v_{i}$ and CH vehicle $v_{j}$. If the vehicle status is in CH, IS or TCH, then the vehicle's BI is set to 0.3. The BI of a vehicle in CM depends on the length of the LET. 

\subsection{Cluster Head Selection Mechanism}

If a vehicle in the IS state receives a hello packet from another vehicle in the IS state or TCH state, this triggers the cluster head selection mechanism. The relative speed and distance can be calculated from the information in the received packet, such as speed, direction, coordinates, and the numeric code of the passing path.

\begin{equation}
\overline{\bigtriangleup v_{i} } = \frac{\sum_{v_{j}\in STS_{i}  }^{}{\left | \bigtriangleup v_{ij}  \right | } }{STS \_ num_{i} } 
    \tag{3}
\end{equation}

\begin{equation}
\overline{\bigtriangleup D_{i} } = \frac{\sum_{v_{j}\in STS_{i}  }^{}{\left | \bigtriangleup D_{ij}  \right | } }{STS\_num_{i} } 
    \tag{4}
\end{equation}

LET mean is calculated as follows: 

\begin{equation}
\overline{\bigtriangleup LET_{i} } = \frac{\sum_{v_{j}\in STS_{i}  }^{}{LET_{ij}  } }{STS\_num_{i} } 
    \tag{5}
\end{equation}

where $STS_{i}$ is the set of vehicle $v_{i}$ neighbor vehicle IS, $STS\_num_{i}$ is the number of vehicle $v_{i}$ neighbors. 

\begin{figure}[htbp]
    \centering
    \includegraphics[width=0.45\textwidth]{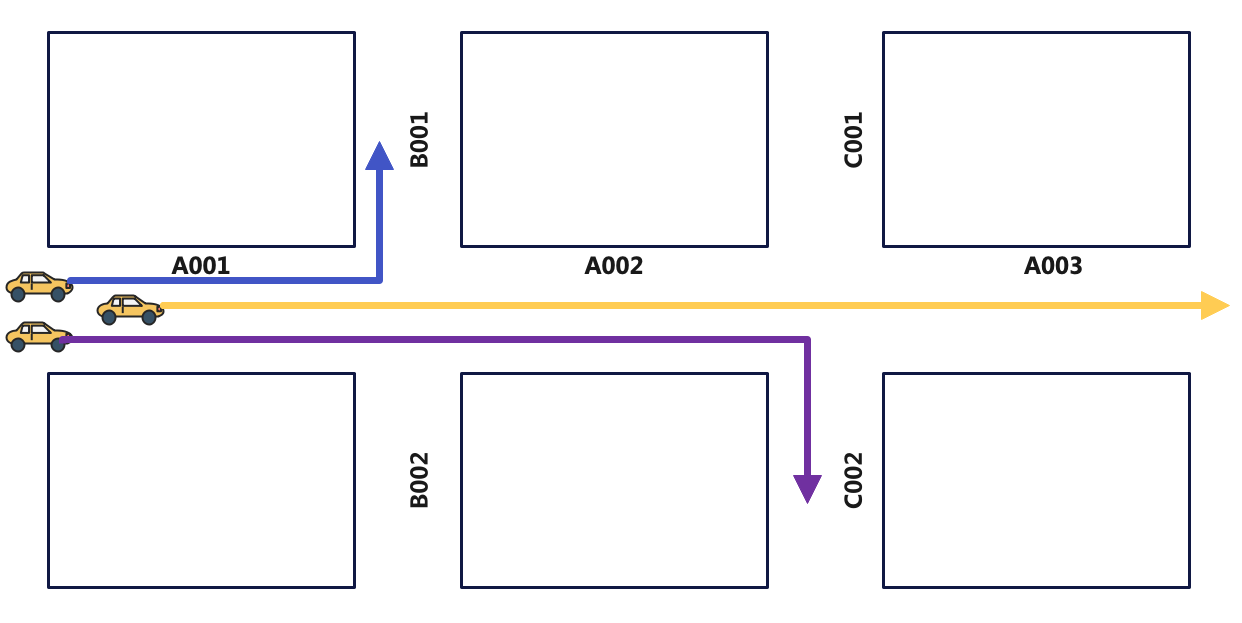}
    \caption{Coding different roads separately and comparing the overlap of vehicle paths}
    \label{fig:2}
\end{figure}

\begin{equation}
\overline{\bigtriangleup PMD_{i} } = \frac{\sum_{v_{j}\in STS_{i}  }^{}{PMD_{ij}  } }{STS\_num_{i} }  
    \tag{6}
\end{equation}
\vspace{+0.08em}
As shown in Fig. 2. pairs in order to plan paths for vehicles, such as online car-hailing and buses. The path code matching degree (PMD) can be calculated from the received hello packets. $PMD_{ij}$ represents the number of overlapping path codes of $v_{i}$ and $v_{j}$, and $PMD_{max}$ is the vehicle with the longest overlapping path code. For vehicles without planned paths, PMD is set to the mean value. 
The $v_{i}$ weighted sum is calculated as follows:
\begin{equation}
\begin{aligned}
MW_{i} = \beta  \times \frac{\overline{\bigtriangleup v_{i} } }{\bigtriangleup v_{max} } &+  \delta  \times \frac{\overline{\bigtriangleup D_{i} } }{\bigtriangleup D_{max} }  +  \varepsilon  \times \left (1 -  \frac{\overline{LET_{i} } }{ LET_{max} } \right )\\&+ \mu  \times \left ( 1- \frac{\overline{PMD_{i} } }{PMD_{max} }  \right )  
\end{aligned}
    \tag{7}
\end{equation}

\begin{equation}
\beta +  \delta +  \varepsilon  +  \mu = 1
    \tag{8}
\end{equation}

Where $MW_{i}$ represents the weighted sum of the migration rates of vehicles $v_{i}$. This value is calculated by considering the average speed, relative distance, link survival time, and overlap of planned paths of the vehicles. Specifically, the $MW_{i}$ of a vehicle is directly proportional to the average speed and relative distance and inversely proportional to the link survival time and planned path overlap. In other words, the smaller value of $MW_{i}$ indicates the higher stability of the node. Therefore, the vehicle with the smallest $MW_{i}$ is found among the neighboring nodes of the IS and is selected as the CH node. The node that becomes CH also broadcasts hello requests regularly, and if the vehicles in the IS state and TCH state meet the requirement, they will join the cluster.

\subsection{Cluster Formation}

We assume that vehicles enter the roadway one by one according to a specific traffic flow and select cluster heads during the clustering process.
Once the vehicles enter the road, they start from the IS state. A hello message is broadcasted with BI every 0.3 seconds, which contains the identifier, speed, direction of movement, and coordinate information. If the vehicle has planned a path, it will also include the path code to reach the destination. In addition, a timer is set for each vehicle as it enters the roadway. When the timer expires, if the vehicle state is still unselected and there are no neighbor nodes or IS state neighbor nodes around. Then, the vehicle will declare itself as a TCH; however, if there are neighboring points in the IS state or three or more TCH nodes within the transmission range, the cluster formation process will be initiated.

The node in the IS state first starts the computation and selects the node with the minimum $MW_{i}$ value as the cluster head. Then, it starts broadcasting the hello message. The node will determine by itself whether it is located within the cluster head's transmission range and traveling in the same direction. If the conditions of transmission range and traveling in the same direction are satisfied, then the node will send an acknowledgment request to the cluster head with its hello packet. At the same time, it will change its state from IS state to CM and set the cluster-ID.This handshake mechanism, which autonomously determines whether to join a cluster based on the receipt of the hello packet, significantly reduces the computational burden of the CHs and makes the formation of clusters more efficient.

In addition, if there exist two CHs entering each other's communication range and the distance L between them is less than the threshold TR, which satisfies the cluster fusion condition, the cluster head with a higher $MW_{i}$ value will give up its cluster head state and become CM state instead. This mechanism helps to avoid excessive cluster head competition and conflict to improve the stability and efficiency of clusters.

\section{Q-learning Cluster-based Routing And CMT Selection Mechanism}

QCBR is a combination of Q-learning and cluster-based routing divided into three levels. First, it clusters the vehicles and generates a list of neighboring clusters; then, in the clustered cluster, it finds the optimal ordering of clusters by the Q-learning algorithm. Finally, it assigns CMT communication nodes based on the position information of the previous and next clusters to realize unicast communication. In the following, we will describe the Q-learning and CMT node selection process in detail.

\subsection{Q-learning}
Typically, a reinforcement learning model chooses an action $a$ to perform in an environment state $s $ and then enters the next state. The agent evaluates the value of the "state-action" based on the reward value fed back from the environment and the next state after acting. If an action $a_{t}$ of an intelligent body brings a positive reward value $R_{t}$, then the action that performs this strategy will be reinforced. The agent's task is to find the optimal strategy to maximize the expected sum of discounted rewards in a discrete state.
Model-based algorithms usually require that the reward is only available when the target node receives a message. Therefore, model-based approaches may be less practical in real-world applications. In contrast, Q-learning is a model-free algorithm that finds the optimal policy through a Markov Decision Process (MDP), even if Agnet has no prior knowledge of the impact on the environment. This approach allows for improved decision-making strategies through continuous learning and experimentation without needing a priori Q-values.

Finding the optimal cluster ordering using reinforcement learning involves the following steps:

\begin{itemize}

\item \textbf{Learning Environment}: Learning environment with clustered clusters as agent.

\item \textbf{State Space}: The state space is a cluster of one or more vehicles.

\item \textbf{Action}: When a CMT node transmits a packet to a CMT node in another cluster, it indicates a change in the packet state. It is also the transmission of a packet from one cluster to another.

\item \textbf{Agent}: Each packet transmitted across a cluster can be viewed as an agent. 

\item \textbf{reward Function}:  The value obtained by an agent performing an action is known as the reward value, and a specific reward value is obtained whenever a packet travels from the current cluster to another.

\end{itemize}

The Q-learning algorithm mainly utilizes  Equation (9) to iteratively update the Q value until convergence. The optimal policy can be constructed by selecting the action with the highest value in each state.

\begin{equation}
 \begin{aligned}
Q\left ( s_{t} ,a_{t}  \right ) &\gets \left ( 1- \alpha  \right ) Q\left ( s_{t},a_{t}   \right ) 
+  \alpha \left ( f_{R}\left ( s_{t},a_{t}   \right )\right.\\
&\left. +  \gamma \underset{a^{'}  }{max } Q\left ( f_{S} \left ( s_{t},a_{t}   \right ) ,a^{'} \right )   \right ) 
 \end{aligned}
    \tag{9}
\end{equation}

In Equation (9), $Q\left ( s_{t},a_{t}   \right )$ is the actual value of the state-to-action correspondence, often referred to as the Q-value. The learning rate $\alpha $ determines to what extent the newly acquired information will overwrite the old information. With a learning rate of 0 , the agent does not learn anything, while with a learning rate of 1 , the agent only considers the latest information. $f_{R} \left ( s_{t},a_{t}   \right ) $ is the reward value, and $\gamma $ is a discount factor determining future rewards' importance. $a^{'}$ denotes the following action that corresponds to the next state.

Since the states and actions are discrete, the reward function is bounded, while the $\left ( s_{t},a_{t}   \right )$ pair can be accessed an infinite number of times, and the Q-learning function converges in a finite amount of time. This means that Q-learning can converge to an optimal Q-value function within a finite number of learning iterations. In the design of QCBR routing, the main goal is to transmit the message from the cluster of the message-sending vehicle to the destination with high probability, which is similar to the typical example in the Q-learning algorithm. In this, the agent learns the Q-value function to make decisions to maximize the cumulative reward. Each cluster head maintains a table of Q-values that accurately represents the relationships between clusters. When the cluster head needs to select the next cluster to deliver a message, it selects the cluster with the highest Q-value as the next hop cluster.

\begin{figure}[htbp]
    \centering
    \includegraphics[width=0.45\textwidth]{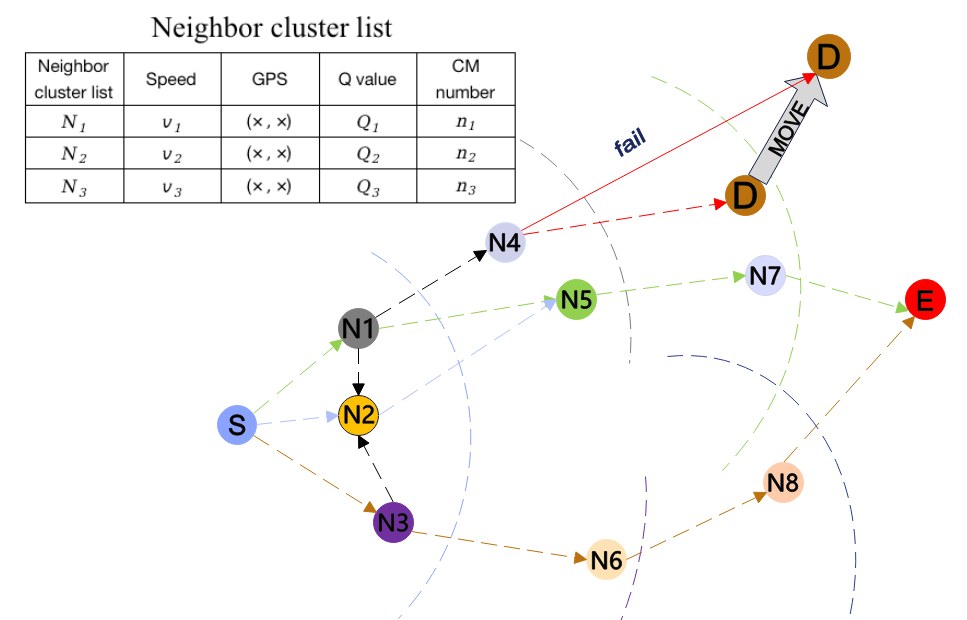}
    \caption{Reinforcement learning process}
    \label{fig:3}
\end{figure}

Fig. 3 shows how Q-learning works, where different dots represent different clusters, and each cluster $N_{i}$ represents a different state in the state space $S$. The arrows indicate the direction of packet transmission, and packet transmission between different clusters has different Q-values. In this example, the transmission operation connecting to the destination node E will be rewarded 100 points, while other operations are not rewarded.
The message is transmitted from the start point S to transmit to the end point E. At the same time, cluster D is about to leave the area, resulting in the inability to efficiently transmit the message through cluster D to the end point E. Therefore, there are three different feasible ways of selecting the clusters that can determine the transmission path of the message: $S\to N_{1} \to N_{5}\to N_{7} \to E$, $S\to N_{2} \to N_{5}\to N_{7} \to E$, $S\to N_{3} \to N_{6}\to N_{8} \to E$. 

\subsection{Cluster-based Routing}

Since the number of vehicles in each cluster may vary, clusters with more members should be considered when selecting the next cluster for message transmission. The advantage of this is that the CH can find an alternative node as the next CMT node faster, even if the CMT node leaves the transmission range. Our goal is to select stable, reliable links and have fewer hops to propagate messages. Therefore, we set the discount factor $\gamma $ as a dynamic parameter positively related to the number of cluster members. We can use a segmented function to represent the discount factor $\gamma $ , where CMN denotes the number of cluster members within a cluster.

\begin{equation}
\overline{CMN} = \frac{1}{N}  {\textstyle \sum_{k= 1}^{N}CMN_{k} } 
    \tag{10}
\end{equation}

Where $\overline{CMN}$ is the cluster membership mean of all clusters in the map and $N$ is the number of clusters in the map.

\begin{equation}
CHV_{ij}  = \sqrt{\left | CHV_{i}^{2} - CHV_{j}^{2}  \right | } 
    \tag{11}
\end{equation}
\begin{equation}
CHD_{ie} \approx \sqrt{\left ( CHX_{i} - CHX_{e}  \right ) ^{2}+ \left ( CHY_{i}- CHY_{e}   \right )  ^{2}  } 
    \tag{12}
\end{equation}

where is the difference in speed between $CHV_{ij}$ cluster heads $CH_{i}$ and $CH_{j}$, and $CHD_{ie}$ is the relative distance between cluster head $CH_{i}$ and end point E. Since the cluster head is always at the center of the cluster, the speed of CH represents the speed of the cluster. Set $CHV_{ij} =  1$ when $CHV_{ij} \le 1$ at that time.

\begin{equation}
\omega = \frac{1}{CHV_{ij} } 
    \tag{13}
\end{equation}
\begin{equation}
\varphi  = \frac{1}{CHD_{ie} } 
    \tag{14}
\end{equation}
\vspace{-0.1em}
\begin{equation}
\gamma = \left\{\begin{matrix} 0.9\omega \varphi  & if & \overline{CMN} \le CMN_{i} \\\\ 0.7\omega \varphi  & if & \frac{\overline{CMN} }{2}\le CMN_{i} < \overline{CMN}  \\\\ 0.5\omega \varphi  & if & 0\le CMN_{i} < \frac{\overline{CMN} }{2} \end{matrix}\right.
    \tag{15}
\end{equation}
\vspace{+0.1em}
Depending on the density of vehicles in different clusters, the $\gamma$ ranges from 0 to 0.9. We can determine whether the density of neighboring clusters is suitable for the next cluster to deliver the message. At the same time, we introduce the relative speed between clusters and the distance from the cluster to the end point E as parameters $\omega$ and $\varphi$ that affect the discount factor. Smaller relative speed between clusters usually corresponds to larger $\gamma$ values. Ultimately, we can use the $\gamma$ value to adjust the routing decision to optimize the message delivery efficiency better.

\begin{algorithm}
\caption{QCBR: Q-learning and Cluster Based Routing}
\begin{algorithmic}[1]
\Statex Vehicle $S$ has a message destined to a specific location $E$
\If{$E$ is within cluster $N_{i}$ transmission range}
  \State Transfer the message to $E$
    \If{Messages are transmitted within clusters}
        \State $TTL = TTL - 0.5$
    \Else
        \State Messages are transferred between clusters
        \State $TTL = TTL - 1$
        \If{$TTL > 0$ and neighbor($N_{i}$)$\neq\emptyset$ }
            \ForAll{neighbor($N_{i}$)}
               \State Calculate the neighbor cluster with 
               \Statex \qquad \qquad \qquad the largest Q
            \EndFor
               \State Select the cluster with the largest Q value
               \Statex \qquad \qquad \quad as the next hop cluster
        \EndIf
    \EndIf
\Else
    \State Discard the message
\EndIf
\end{algorithmic}
\end{algorithm}

As shown in Algorithm 1, we assume that each vehicle has an offline-learned Q-value table that follows Equation (9). In addition, we set a $TTL$ field for each packet to determine the survival time of the message. In each transmission, $TTL-0.5$ is for intra-cluster node transmission, and $TTL-1$ is for inter-cluster transmission. Suppose the message cannot reach the destination directly. If $TTL>0$, the message will continue transmitting according to the specific policy. Once $TTL\le 0$, the message will be discarded.

\subsection{Transmission Relay Node CMT Selection Mechanism}

To achieve unicast communication after clustering, specific cluster members must be selected as CMT nodes. However, some CM nodes may be about to leave the cluster, and selecting these soon-to-be-leaving CM nodes as CMT nodes will result in frequent disconnections and reconnections of the communication link. This will adversely affect the end-to-end communication delay and data delivery rate. Therefore, we have designed a selection mechanism for CMT nodes, the details of which are given below.

For those vehicles that have planned their paths, when $PMD_{ij}  = 0 , 1$ , it means that $v_{i}$ and $v_{j}$ do not have overlapping road segments or have only one overlapping path, which means that they are about to leave the communication range of the current cluster head, and therefore are labeled as unstable nodes (UN) .
\begin{equation}
PMD_{ij} = \left\{\begin{matrix}  0,1& unstable\\  \ge 2& stable\end{matrix}\right.
    \tag{16}
\end{equation}

\begin{figure}[htbp]
    \centering
    \includegraphics[width=0.45\textwidth]{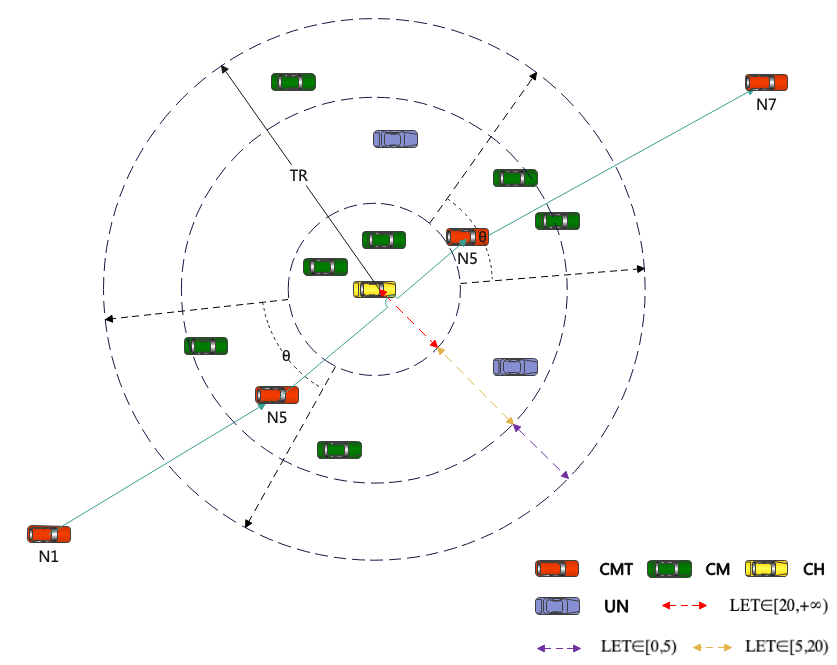}
    \caption{Selection of CMT nodes within a cluster}
    \label{fig:4}
\end{figure}
\vspace{-0.2em}
As shown in Fig. 4. When the previous cluster CMT node sends a message to pass the request, the cluster head selects the node in the cluster that is located within the $LET\in \left [ 0,20 \right )$ in the range of the pinch angle $\theta$ of the aligned node based on the coordinates of the previous cluster CMT node and its LET value. The CMT node selection is categorized into the following scenarios:
\vspace{-0.2em}
\begin{itemize}

\item \textbf{Case 1:} First, the CM with $LET\in \left [ 0,20 \right )$ in the range of pinch angle $\theta$ and the smallest value of $MW_{i}$ is selected as the CMT node of the cluster. Because the nodes in this range are located in the middle of the cluster, a balance is achieved between node stability and hop count.
\item \textbf{Case 2:} Then, the node with $LET\in \left [ 0,5 \right )$ and the smallest $MW_{i}$ value is selected in the range of angle $\theta$ . The nodes in this range are usually located at the edge position of the cluster and can directly receive information from the CMT nodes in the previous cluster, but due to the small value of LET, there may be a risk of disengaging from the cluster, which can lead to link fluctuation.
\item \textbf{Case 3:} Next, the node with the smallest value of $MW_{i}$ in $LET\in \left [ 20,+\infty  \right )$  is selected because the node in this position is relatively close to the CH, and there is no need to consider the pinch range. Although these nodes are more stable and less likely to fall out of the transmission range of the cluster, they are usually farther away from the CMT node of the previous cluster, which may lead to an increase in the number of hops for data transmission.
\item \textbf{Case 4:} Finally, when the vehicle density is too low to fulfill the previous conditions, UN nodes or CH nodes are selected as CMT nodes.

\end{itemize}

 When a CMT node within a cluster loses connectivity, the CMT node selection mechanism is retriggered. The cluster alignment formed by Q-learning informs the cluster head about the next cluster that needs to transmit information. Then, based on the cluster head's list of neighboring clusters, it determines the location of the next cluster head and its distance L. When the CMT node $LET\in \left [ 0,5 \right )$ is in the cluster, $LET\in \left [ 20,+\infty  \right )$ is selected as the next CMT node. If the cluster head spacing L satisfies $TR\le L< 2TR$ and the CMT node $LET\in \left [ 5,+\infty  \right )$ within the cluster, then the information transmission requests directly to the next cluster. Suppose $2TR< L$, the cluster head is triggered to select another intra-cluster CMT node. The selection step is the same as above. When a TCH node receives a transmission request, it automatically changes to a CMT state. This mechanism ensures message delivery in case of low vehicle density.

 \subsection{QCBR Protocol Message Forwarding Process}
 
This protocol has been designed considering different vehicle densities in the city. Usually, QCBR selects clusters with higher vehicle density to disseminate messages. However, in some cases, such as suburban areas and other environments with lower vehicle densities, vehicles may not be able to form clusters, which results in messages not being transmitted. Therefore, we introduce a timer mechanism for vehicles in the IS state. When the timer expires, if there is no other node in the IS state around, then the state of the vehicle will be changed to TCH, thus enabling it to participate in message transmission.

\begin{figure}[htbp]
    \centering
    \includegraphics[width=0.45\textwidth]{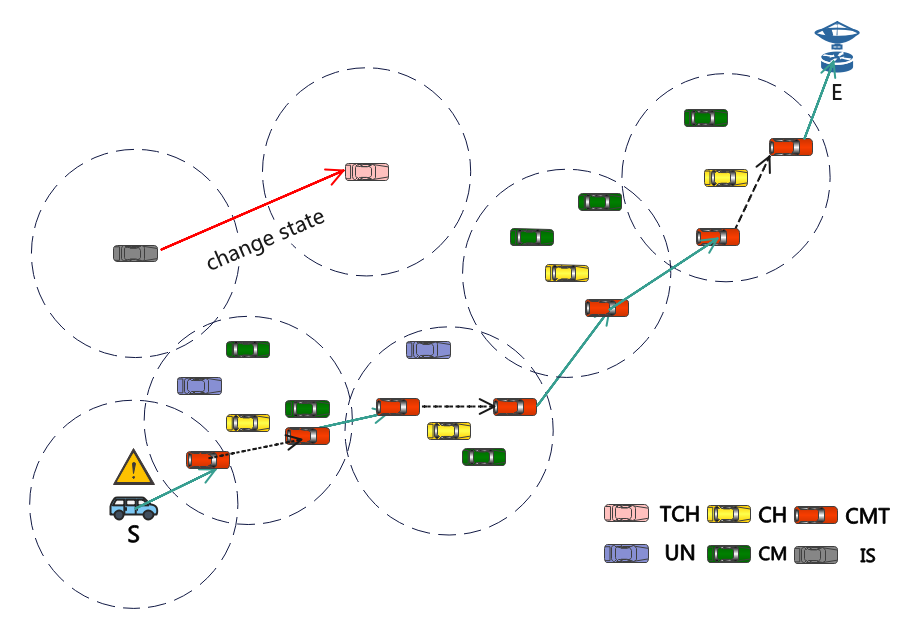}
    \caption{Example of message forwarding with QCBR}
    \label{fig:5}
\end{figure}

As shown in Fig. 5, the message is sent from the blue vehicle S node, forwarded by the CMT node carrying the message, and finally reaches the destination node E.

\section{SIMULATIONS}
\begin{figure*}[htbp]
    \centering
    \begin{subfigure}[b]{0.3\textwidth}
        \includegraphics[width=\textwidth]{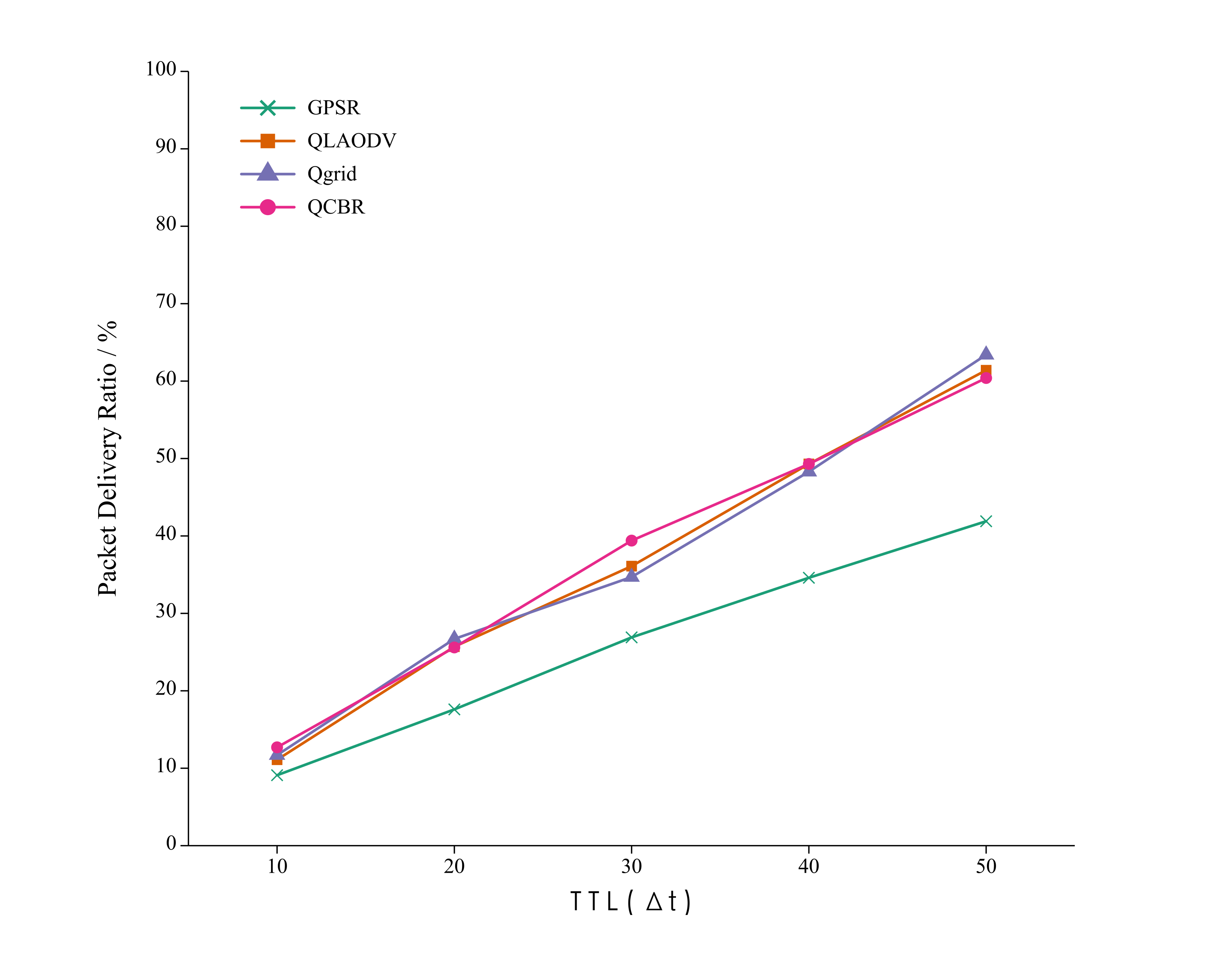}
        \caption{Packet Delivery Ratio}
    \end{subfigure}
    \hfill
    \begin{subfigure}[b]{0.3\textwidth}
        \includegraphics[width=\textwidth]{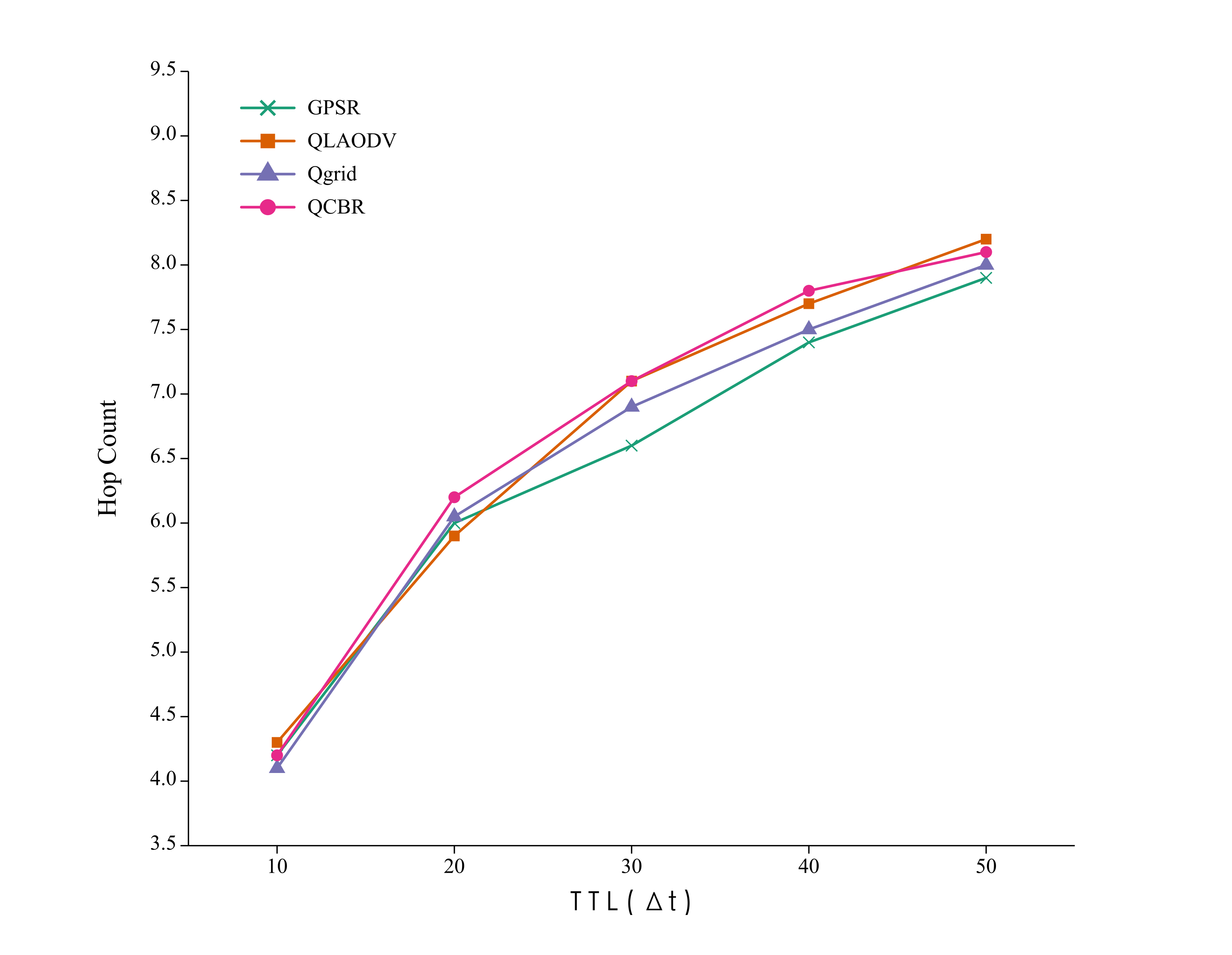}
        \caption{Hop Count}
    \end{subfigure}
    \hfill
    \begin{subfigure}[b]{0.3\textwidth}
        \captionsetup{skip=0pt} 
        \includegraphics[width=\textwidth]{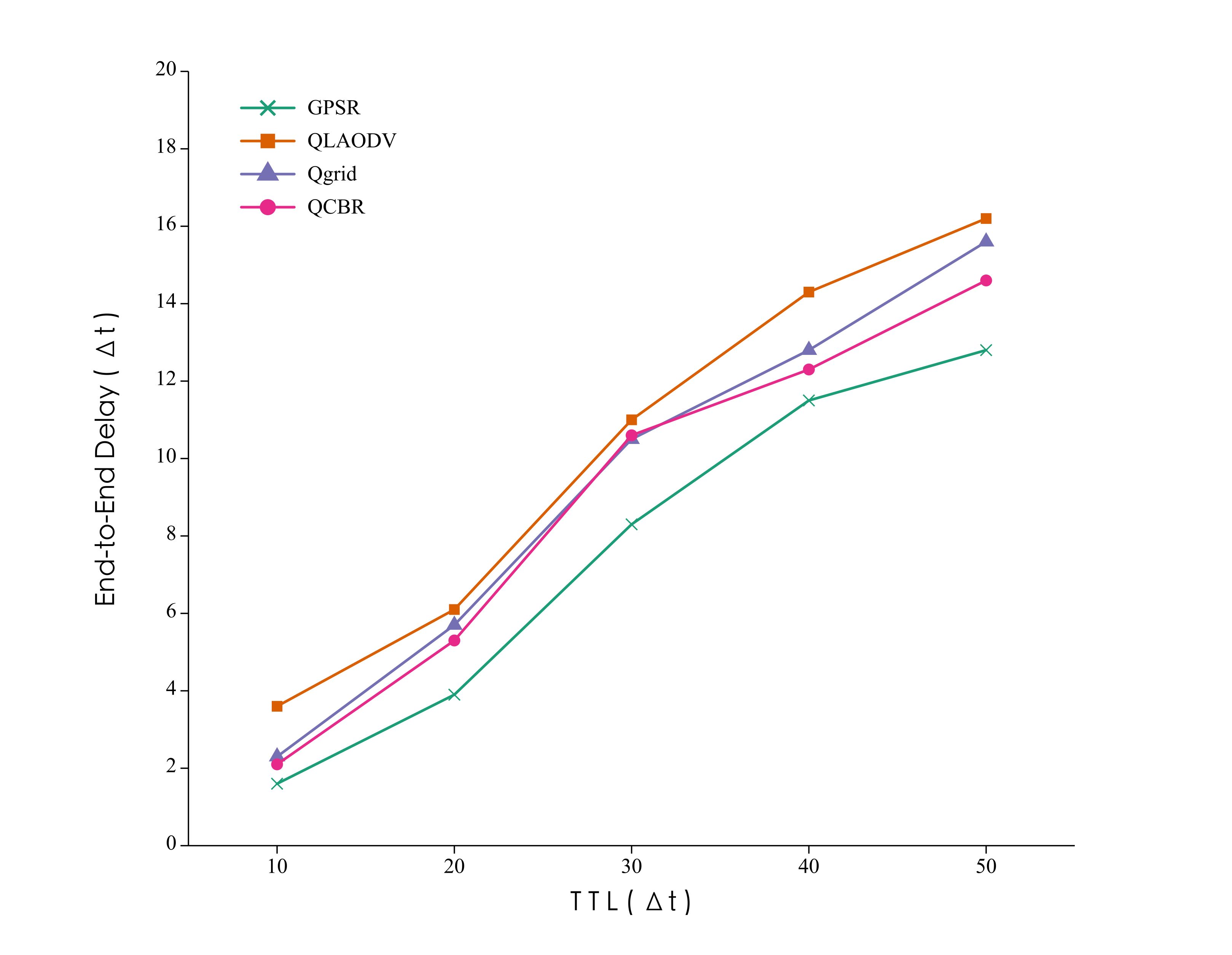}
        \caption{End-to-End Delay}
    \end{subfigure}
    \caption{Comparison of Routing Simulations for Different Protocols (vehicle node set to 200, PMD set to 40)}
\end{figure*}
\begin{figure*}[htbp]
    \centering
    \begin{subfigure}[b]{0.3\textwidth}
        \includegraphics[width=\textwidth]{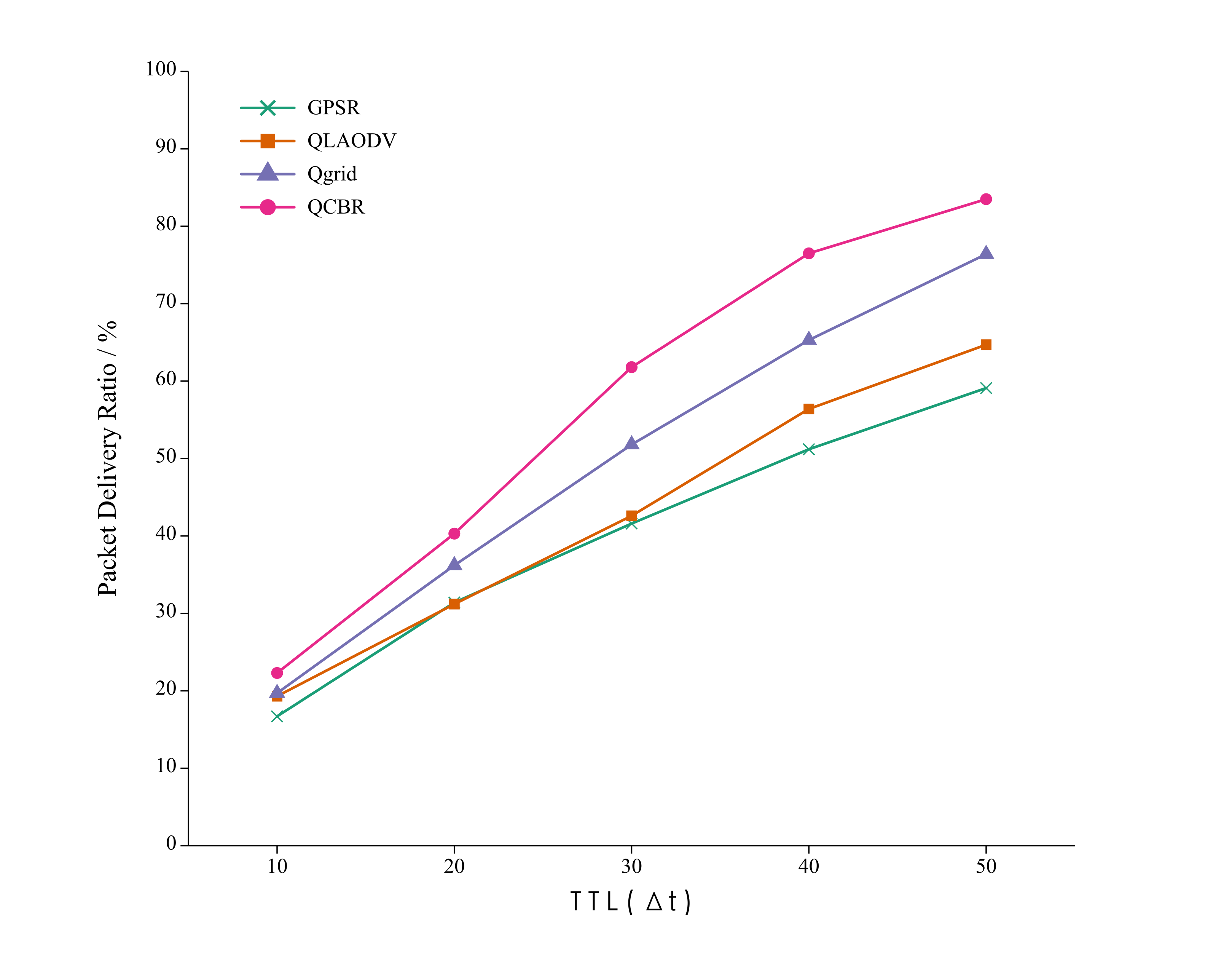}
        \caption{Packet Delivery Ratio}
    \end{subfigure}
    \hfill
    \begin{subfigure}[b]{0.3\textwidth}
        \includegraphics[width=\textwidth]{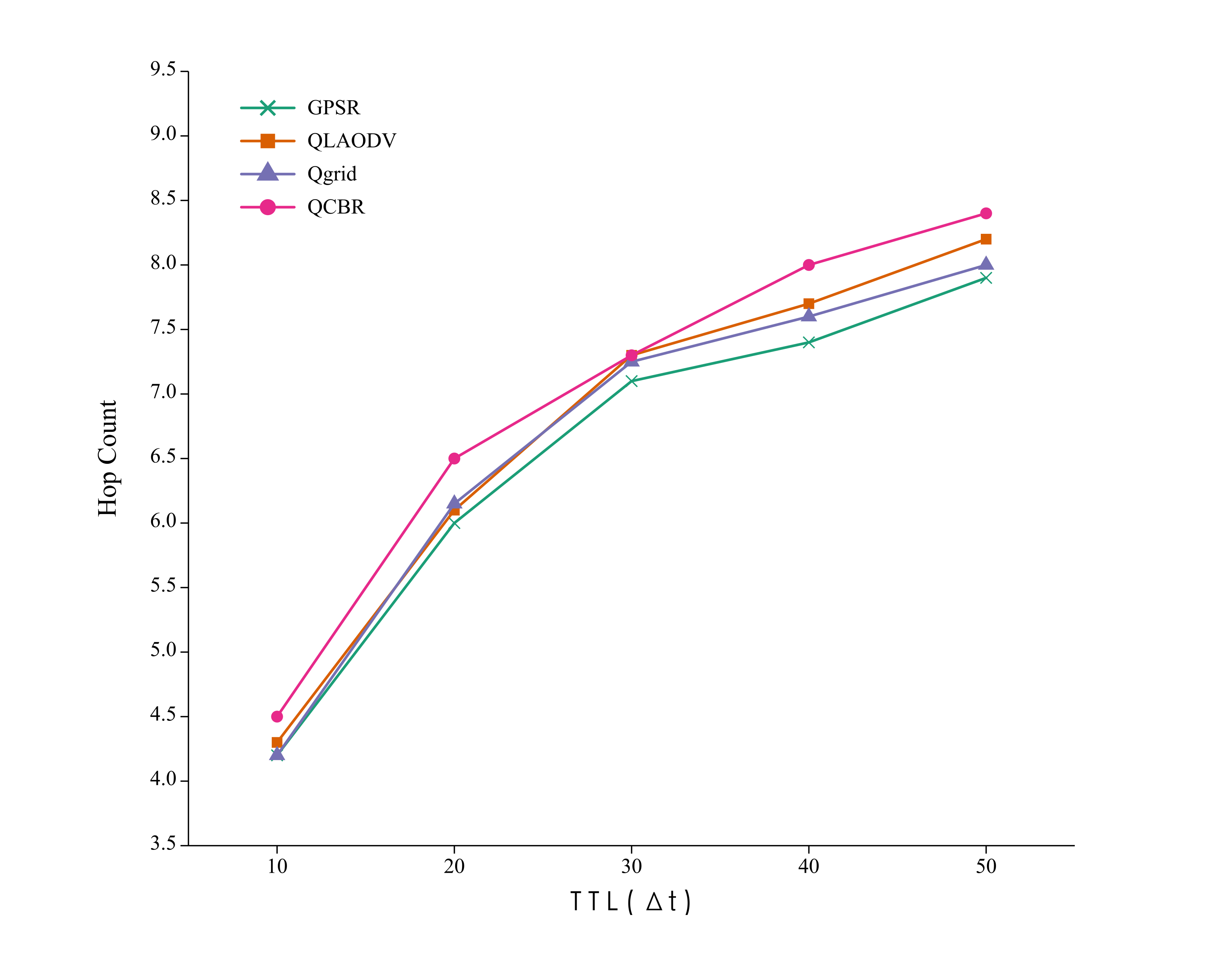}
        \caption{Hop Count}
    \end{subfigure}
    \hfill
    \begin{subfigure}[b]{0.3\textwidth}
        \captionsetup{skip=0pt} 
        \includegraphics[width=\textwidth]{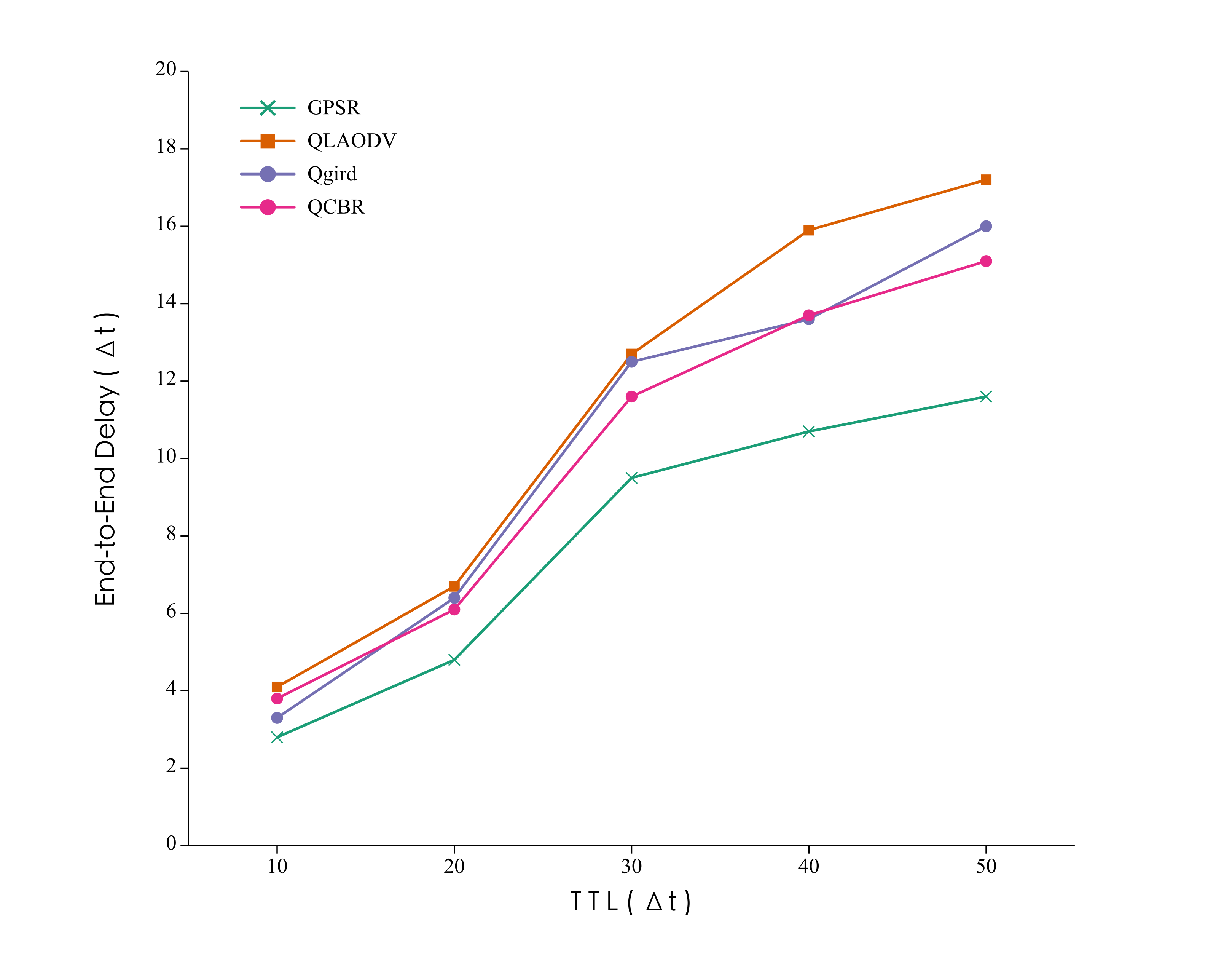}
        \caption{End-to-End Delay}
    \end{subfigure}
    \caption{Comparison of Routing Simulations for Different Protocols (vehicle node set to 400, PMD set to 80)}
\end{figure*}
\begin{figure*}[htbp]
    \centering
    \begin{subfigure}[b]{0.3\textwidth}
        \includegraphics[width=\textwidth]{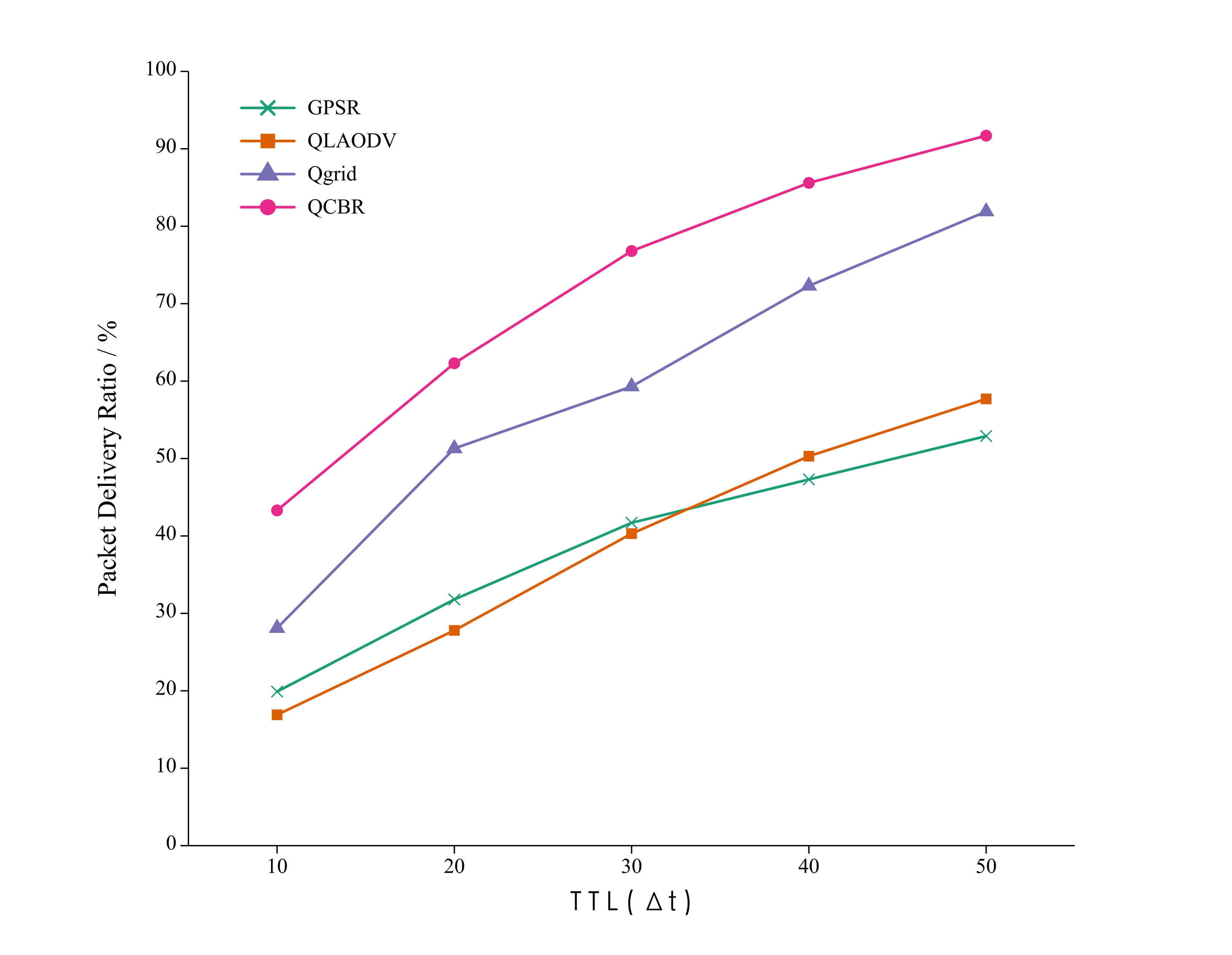}
        \caption{Packet Delivery Ratio}
    \end{subfigure}
    \hfill
    \begin{subfigure}[b]{0.3\textwidth}
        \includegraphics[width=\textwidth]{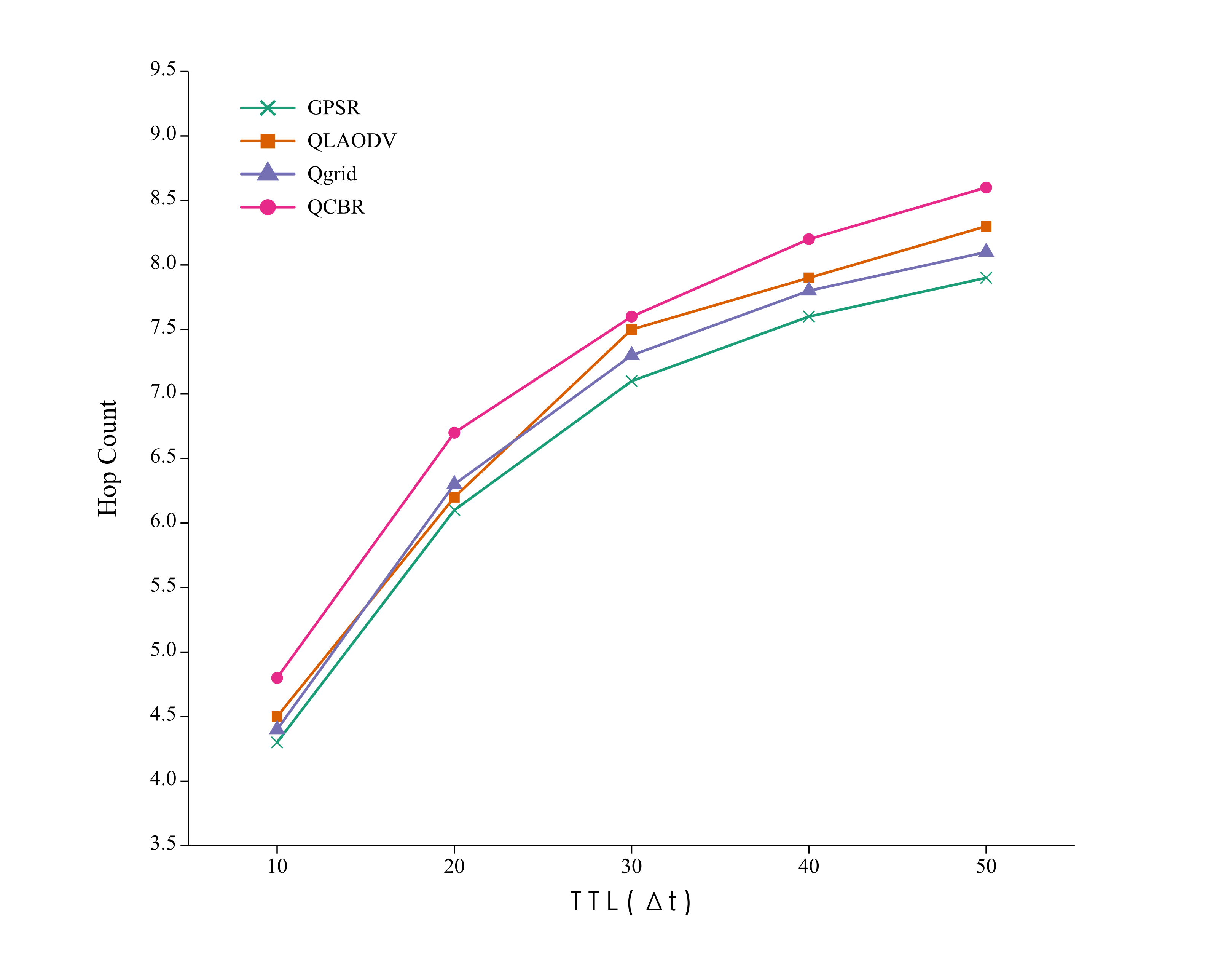}
        \caption{Hop Count}
    \end{subfigure}
    \hfill
    \begin{subfigure}[b]{0.3\textwidth}
        \captionsetup{skip=0pt} 
        \includegraphics[width=\textwidth]{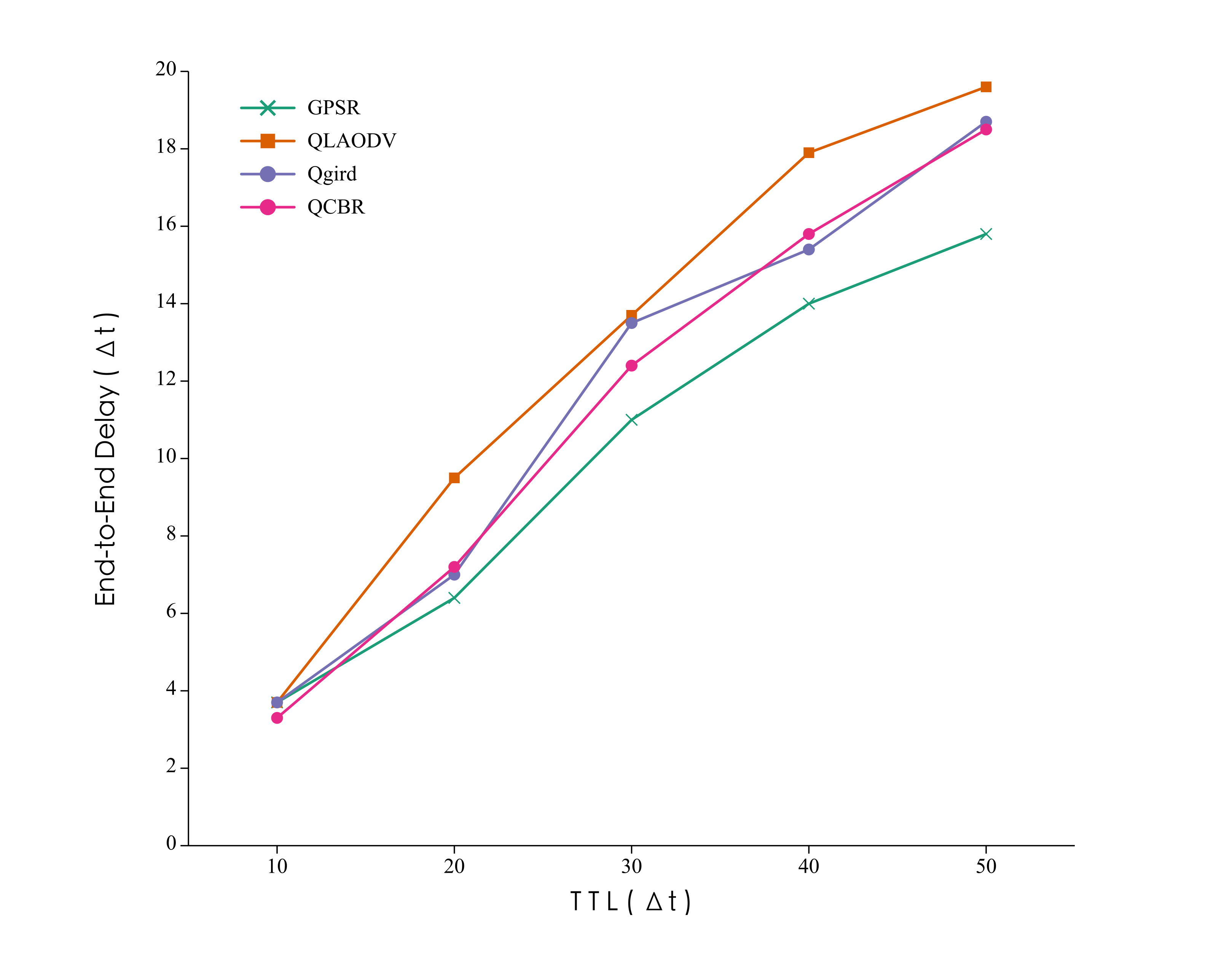}
        \caption{routing cost}
    \end{subfigure}
    \caption{Comparison of Routing Simulations for Different Protocols (vehicle node set to 600, PMD set to 120)}
\end{figure*}

We will validate the performance of QCBR in different scenarios and compare it with four other existing location-based, reinforcement learning, and geo-grid-based protocols. The detailed setup and experimental results are shown below.

\subsection{Experimental Scenarios and Assessment Indicators}

To simulate a realistic traffic environment, we intercepted a 3000m $\times$ 3000m area map using Open Street Map, which contains a complex road network covering both dense and sparse road areas. We use sumo, omnet++, and veins for simulation experiments. The road simulation tool sumo (Simulation of Urban Mobility) handles large road networks and simulates traffic models. omnet++ is a discrete event simulator based on C++. Veins is a simulation framework for inter-vehicle communication that consists of an event-based network simulator and a road simulation. Network simulator and road simulation model. The clustering algorithm and Q-learning settings during the simulation are shown in the following table.

\begin{table}[htbp]
    \centering
    \caption{Simulation setup}
\begin{tabular}{cc}
   \toprule
   Parameter & Value or Range \\
   \midrule
   Simulation scene & 3000m $\times$ 3000m \\
   $\alpha$ & 0.8 \\
   $\omega$ & $\left ( 0,1 \right ] $ \\
   $\varphi$ & $\left ( 0,1 \right ) $ \\
   $\gamma$ & $\left ( 0,0.9 \right ) $ \\
   CH broadcast range & 260 \\
   TR & 200 \\
   Data rate & 10 packet/s \\
   $TTL$ & 10, 20, 30, 40, 50 \\
   Number of Vehicles & 200, 400, 600 \\
   PMD & 40, 80, 120 \\
   reward R & 0, 100 \\
   vehicle speed & 5m/s~20m/s \\
   LET & $\left [ 0,+\infty  \right ) $ \\
   $\theta$ & $90^{\circ}$ \\
   \bottomrule
\end{tabular}
\end{table}

We evaluated the experiment using the following metrics:
\begin{itemize}
\item \textbf{Hop Count}: The average number of hops per successful packet transmission.
\item \textbf{Packet Delivery Ratio}: Ratio of successfully transmitted packets to the destination to the total number of packets sent by the source node.
\item \textbf{End-to-End Delay}: The average time experienced by a packet to be sent from the source node to the final destination. 
\end{itemize}

\subsection{Comparing Routing Protocols}
We use the following protocols for comparison with our proposed QCBR:
\begin{itemize}

\item \textbf{GPSR}:GPSR is a geographic location-based greedy routing algorithm that selects the nearest neighboring node to the target node as the next hop. The packets are transmitted through a series of peripheral nodes.
\item \textbf{QLAODV}: OLAODV is a routing protocol based on a Q-learning approach where each vehicle maintains a Q-value table and selects the neighbor with the largest Q-value as the next hop node.
\item \textbf{Qgrid}: Qgrid is a geographic grid-based routing protocol that assumes a digital map for each vehicle. It divides the map into grids of uniform size, finds the best geographic grid transmission path through reinforcement learning, and unicasts the signal along the grid while guiding the vehicle along the best path. In the Qgrid setup, the reinforcement learning parameter part is consistent with ours, and the grid length is set to 200m.

\end{itemize}
\subsection{Analysis of Experimental Results}

In urban environments, vehicle density is not uniformly distributed. Based on this, we set different vehicle node densities (200, 400, 600) to verify the transmission effect under different vehicle densities. Meanwhile, we set some vehicles as PMD nodes with known paths corresponding to buses and internet taxis traveling along fixed routes or planned paths in urban environments. The number of densities corresponding to one-fifth of PMDs to all the vehicles are (40, 80, 120).

In our experiments, we compare the performance differences under different vehicle densities. As shown in Fig. 6(a), the reinforcement learning-based algorithm significantly outperforms the location-based GPSR algorithm regarding data delivery rate at low vehicle densities. This is because the reinforcement learning algorithm considers the stability of neighboring nodes and does not just look for the node closest to the destination. However, since the increase in hops slightly increases the end-to-end delay, as shown in Fig. 6(c), the end-to-end delay is slightly higher than GPSR, but the data delivery rate is substantially higher. Our algorithm does not draw a significant gap with Qgird in the low-density setting. There are mutual advantages and disadvantages at different TTLs. Slightly higher than QLAODV in end-to-end data latency.

In the medium vehicle density condition, as shown in Fig. 7(a), we can see that our data delivery rate is improved. In case of increased vehicle density, CMT will have more alternative nodes. This improves the link stability. The effect improvement is relatively significant compared to QLAODV and GPSR. Compared to the grid-based reinforcement learning routing algorithm Qgird, our data delivery rate is slightly improved; there are certain limitations in the grid-based routing of Qgird. The division of the grid size significantly impacts the results; the division of the grid size is too large. However, it can speed up the convergence. Still, the grid has too many vehicles, and using the greedy algorithm to select the transmission nodes makes the reinforcement learning effect not obvious. As in Fig. 7(b), we observe that QCBR and Qgird have higher hop counts and delays, mainly due to selecting the likely successful paths rather than the shortest path.

In the case of high-density vehicles, QCBR significantly improves the data delivery rate compared to other routing protocols. QCBR introduces the mechanism of computationally planning paths, making distinguishing unstable nodes easier. From 8(a), it can be seen that the data delivery rate of GPSR and QLAODV even decreases when the vehicle density is high; GPSR with more vehicles causes the road situation to become complex. Finding the nearest node to the destination is more complex, thus affecting the data delivery rate. QLAODV routing state space becomes larger when there are too many neighboring nodes, leading to difficulty in convergence. In contrast to 7(a), more vehicle nodes are needed to reduce the data delivery rate.8 (c)The results show a significant increase in the delay of GPSR and QLAODV compared to the low-density case.

\section{CONCLUSION}

We propose a unicast routing protocol called QCBR, which addresses the challenge of slow convergence of reinforcement learning in VANET environments. By combining reinforcement learning and clustering algorithms, the QCBR protocol significantly improves the data reception rate of vehicles reaching their destinations. Even in the case of unicast link disconnection, the QCBR protocol can quickly select alternative vehicles as CMT nodes to ensure communication continuity. In addition, our protocol exhibits stable communication performance in environments with different vehicle densities. Through a series of simulation experiments, we demonstrate that the QCBR protocol improves the data delivery rate and reduces the end-to-end delay relative to other unicast routing protocols based on reinforcement learning.

\vspace{12pt}

\end{document}